\documentclass[11pt]{article}
\usepackage{fullpage}

\usepackage{graphicx}
\usepackage{amsmath}
\usepackage{amssymb}
\usepackage{amsthm}
\usepackage{algorithm}

\usepackage{paralist}

\usepackage{color}
\usepackage[pagebackref]{hyperref}
\hypersetup{colorlinks=true,citecolor=red,linkcolor=blue}

\newtheorem{theorem}{Theorem}[section]
\newtheorem{prop}[theorem]{Proposition}

\newtheorem{fact}[theorem]{Fact}

\newtheorem{question}[theorem]{Question}
\newtheorem{lemma}[theorem]{Lemma}
\newtheorem{definition}[theorem]{Definition}

\newcommand\R{\mathbb{R}}

\newcommand\X{{\cal X}}
\newcommand\Y{{\cal Y}}

\newcommand{\rank}{\operatorname{rank}}

\renewcommand{\tilde}{\widetilde}
\renewcommand{\hat}{\widehat}
\newcommand{\wt}{\widetilde}

\renewcommand\d{\mathrm{d}}

\newcommand{\todo}[1]{{\bf \color{red}}} \newcommand{\tim}[1]{{\color{darkgreen}[Tim: #1]}}

\newcommand{\Zhao}[1]{{\color{red}[Zhao: #1]}}

\usepackage[normalem]{ulem}

\title{Metric Transforms and Low Rank Matrices via Representation Theory of the Real Hyperrectangle}

\date{}

\author{
    Josh Alman\thanks{\texttt{josh@cs.columbia.edu}. Columbia University.}
    \and 
    Timothy Chu\thanks{\texttt{timothyzchu@gmail.com}. Carnegie Mellon University.}
    \and 
    Gary Miller\thanks{\texttt{gm2f@andrew.cmu.edu}. Carnegie Mellon University.}
    \and 
    Shyam Narayanan\thanks{\texttt{shyamsn@mit.edu}. MIT.}
    \and 
    Mark Sellke\thanks{\texttt{msellke@stanford.edu}. Stanford University.} 
    \and 
    Zhao Song\thanks{\texttt{magic.linuxkde@gmail.com}. Princeton University and Institute for Advanced Study.}
}

\begin{document}

\newcommand{\vectorize}[1]{\textsc{Vectorize}\left[ #1 \right]}

\newcommand{\tc}[1]{\noindent{\textcolor{red}{\textbf{\#\#\# TC:} \textsf{#1} \#\#\#}}} 
\newcommand{\sn}[1]{\noindent{\textcolor{blue}{\textbf{\#\#\# SN:} \textsf{#1} \#\#\#}}}
\definecolor{darkgreen}{rgb}{0.0, 0.5, 0.0}
\newcommand{\ms}[1]{\noindent{\textcolor{darkgreen}{\textbf{\#\#\# MS:} \textsf{#1} \#\#\#}}}

\begin{titlepage}
  \maketitle
  \begin{abstract}

In this paper, we develop a new technique which we call representation theory of the real hyperrectangle, which describes how to compute the eigenvectors and eigenvalues of certain matrices arising from hyperrectangles.
We show that these matrices arise naturally when analyzing a number of different algorithmic tasks such as kernel methods, neural network training, natural language processing, and the design of algorithms using the polynomial method. 
We then use our new technique along with these connections to prove several new structural results in these areas, including:
\begin{enumerate}
    \item A function is a positive definite Manhattan kernel if and only if it is a completely monotone function. These kernels are widely used across machine learning; one example is the Laplace kernel which is widely used in machine learning for chemistry.
    \item A function transforms Manhattan distances to Manhattan distances if and only if it is a Bernstein function. This completes the theory of Manhattan to Manhattan metric transforms initiated by Assouad in 1980.
    \item A function applied entry-wise to any square matrix of rank $r$ always results in a matrix of rank  $< 2^{r-1}$ if and only if it is a polynomial of sufficiently low degree. This gives a converse to a key lemma used by the polynomial method in algorithm design.
\end{enumerate} 

 Our work includes a sophisticated combination of techniques from different fields, including metric embeddings, the polynomial method, and group representation theory.

\end{abstract}
  \thispagestyle{empty}
\end{titlepage}

\section{Introduction} 

In this paper, we introduce a new analytic technique we call `representation theory of the real hyperrectangle'. At a high level, this technique gives simple expressions for computing the eigenvectors and eigenvalues of a large class of matrices which are defined in terms of hyperrectangles (high-dimensional analogues of rectangles). We will see that this class of matrices arises frequently in the study of linear algebraic tools for modern machine learning and algorithm design. As a result, we use our new technique to prove a number of new structural results in these areas.

Before getting into the representation theory of the real hyperrectangle in more detail, we first describe our three main applications. First, in Section~\ref{sec:kernel}, we give a classification of positive definite kernels with Manhattan distance input. Second, in Section~\ref{sec:metric}, we categorize all functions which transform Manhattan distances to Manhattan distances or squared Euclidean distances. Third, in Section~\ref{sec:polymeth}, we prove that the only functions which always yield a low-rank matrix when applied entry-wise to a low-rank matrix are low-degree polynomials; this is a converse of a key idea behind the polynomial method in algorithm design and in the training of transformers in natural language processing. Afterwards, in Section~\ref{sec:techniques}, we describe our new tool, representation theory of the real hyperrectangle, and how we use to to yield these applications.

\subsection{Kernel Methods}\label{sec:kernel}
Our first application is to the study of kernel methods in machine learning. Much of the prior work on kernels methods focuses in the Euclidean distance setting. Our new application shows how to classify kernels in the Manhattan distance setting.

We start with defining positive definite kernel under Euclidean space.
 \begin{definition}[Positive definite Euclidean kernel] \label{def:euclineankernel}A function $f$ is a positive definite Euclidean kernel
 if, for any $x_1, \ldots x_n \in \R^d$ for any $n$ and $d$, the matrix $M \in \R^{n \times n}$ with
 
 \[ M_{i,j} = f(\|x_i - x_j\|_2) \]
 is positive semi-definite. Equivalently, $f$ is a positive definite Euclidean kernel if and only if there exists a function $F:\R^d \to \mathcal{H}$\footnote{$\mathcal{H}$ represents Hilbert space.} such that:
 
 \[\langle F(x), F(y) \rangle = f(\|x-y\|_2)\]
 for all $x, y \in \R^d$ for all $d$.
 \end{definition}
 The proof of the equivalence can be found in~\cite{s42}. Positive definite kernels are used in machine learning to separate data embedded in $\R^d$ using linear separator techniques, when the initial data is not linearly separable~\cite{s96, sow01, ss01}. In other words, a positive definite kernel can map points in $\R^d$ which are not linearly separable, to points in potentially higher dimensions which are linearly separable. Finding such an embedding is not an easy task in general, but kernel methods solve this problem~\cite{s96, sow01, ss01}. The key idea is to pick a function $f$ based on the application so that a function $F$ like the one in Definition~\ref{def:euclineankernel} can be found which maps the data points to vectors of possible higher dimensions, after which linear separation can be performed efficiently on these higher dimensional points.
 
Interestingly, linear separator algorithms such as the widely used \emph{Support Vector Machines} (SVMs)~\cite{cv95} can separate the data efficiently as long as $\langle F(x), F(y) \rangle$ is easily computed for any $x, y \in \R^d$, even if $F$ itself cannot be easily computed. By definition of the positive-definite kernel $f$, we know that $\langle F(x), F(y) \rangle = f(\|x - y\|_2)$, which allows us to compute $\langle F(x), F(y) \rangle$ quickly by instead computing $f(\|x - y\|_2)$. In other words, in order to apply linear separator algorithms, it suffices to know that a $F$ \emph{exists}, and not necessarily know what it is or how to compute it.

The core result behind kernel methods is a full classification of all positive-definite Euclidean kernels, showing that a function $f : \R \to \R$ is a positive-definite Euclidean kernel if and only if $f(\sqrt{x})$ is a completely monotone function~\cite{s42, sow01}:  

\begin{definition}[Completely monotone functions~\cite{b29}]\label{def:cm}
A function $f:\mathbb R^+\to\mathbb R_{\geq 0}$ is completely monotone if 
  \[(-1)^k f^{(k)}(x) \geq 0\] 
  for all $k \geq 0, x > 0$. A function $f: \mathbb R_{\geq 0}\to \mathbb R_{\geq 0}$ is completely monotone if $f(0)\geq \lim_{x \to 0^{+}} f(x)$ and $f|_{\mathbb R^+}$ is completely monotone.
\end{definition}
An example of a completely monotone function is $f(x) = e^{-x}$. 
\begin{theorem}[Classification of all positive definite Euclidean kernels~\cite{s42, sow01}]\label{fact:kernel-euc}
Function $f: \R \to \R$ is a positive-definite Euclidean kernel (Definition~\ref{def:euclineankernel}) if and only if $f(\sqrt{x})$ is a completely monotone function.
\end{theorem}
This theorem gives a simple criterion to test whether \emph{any} given function is a positive-definite Euclidean kernel. One famous example of these kernels include The Gaussian kernel: $f_{\sigma}(x) = e^{-\sigma x^2}$ for $\sigma > 0$. Another famous example is called neural tangent kernel: $f(x) = ( \frac{1}{2}- \frac{1 }{2\pi} \arccos ( \frac{1}{2} - \frac{1}{2} x^2 ) ) \cdot (\frac{1}{2} - \frac{1}{2}x^2)$\footnote{This equation is corresponding to the ReLU activation function. For other activation functions, the equation will be different.}, which recently proposed by machine learning community \cite{jgh18} and it plays a crucial role in showing the convergence of deep neural networks with non-linear activation functions \cite{ll18,dzps19,als19_dnn,als19_rnn,sy19,lsswy20,bpsw21}. This theory allows practitioners to describe all positive definite Euclidean kernels.

\subsubsection{Main Result}

In our paper, we classify all positive-definite Manhattan kernels. These kernels are widely used in machine learning for physical and chemical  applications~\cite{flla15, l18, lrrk15}. A notable example of such a kernel is the Laplace kernel $f_\sigma (x) = e^{-\sigma x}$ which is commonly used in classification tasks~\cite{bmm18}. However, a full description of all positive-definite Manhattan kernels  was not known before our work.
 \begin{definition}[Positive definite Manhattan kernel]\label{def:manhattan_kernel} 
 A function $f$ is a \emph{positive definite Manhattan kernel} if, for any $x_1, \ldots x_n \in \R^d$ for any $n$ and $d$, the $n \times n$ matrix $M$ with
 \begin{align*}
  M_{i,j} = f(\|x_i - x_j\|_1) 
 \end{align*}
 is positive semi-definite. 
 \end{definition}
 
 Our main result is as follows:
\begin{theorem}[Main result, informal statement of Theorem~\ref{thm:formal_kernel_manhattan}] \label{thm:informal_kernel_manhattan} 
$f$ is a positive definite Manhattan kernel (Definition~\ref{def:manhattan_kernel}) if only if $f$ is completely monotone (Definition~\ref{def:cm}).
\end{theorem}

Theorem~\ref{thm:informal_kernel_manhattan} classifies all positive-definite kernels when the input distance is Manhattan. It was previously known that completely monotone functions are positive definite Manhattan kernels~\cite{s38, a80}, but it was not known these were the only such functions. Interestingly, our new classification is similar to Theorem~\ref{fact:kernel-euc}, but without a square root applied to the input. Prior to our result, one could have imagined that there are other positive definite Manhattan kernels to use in SVMs than were previously known. However, our result shows that there are no other such kernels.

\subsection{Metric Transforms} \label{sec:metric}

Our second application is to \emph{metric transforms}, a mathematical notion introduced by Schoenberg and Von Neumann~\cite{ns41}. 

\begin{definition}[Metric transform]\label{def:metric-transform}
Suppose $\X$ and $\Y$ are semi-metric spaces\footnote{A semi-metric satisfies all the axioms for a metric except possibly the triangle inequality; the square of the Euclidean distance gives rise to a semi-metric.}. Function $f$ \textbf{transforms} $\X$ to $\Y$ if, for any finite set $S \subseteq \X$, there is a function $F: \X \to \Y$ such that
\[f(d_{{\cal X}}(x_1,x_2)) = d_{{\cal Y}}(F(x_1), F(x_2)) ,
\]
for all $x_1, x_2 \in S$.
 \end{definition}
 Metric transforms arise naturally in many settings where one wants to transform a set of points from a metric space while maintaining some of the metric structure between them. They have proven useful in many areas including 
 sketching and embedding norms~\cite{akr15}, algorithms to compute a manifold geodesic~\cite{cms20}, machine learning~\cite{o96, ssb+97}, harmonic analysis~\cite{a50,lllh18,kw71}, complex analysis~\cite{a50}, and PDE theory~\cite{fs98, cfw12}. Typically we have particular metric spaces $\X$ and $\Y$ of interest, as well as certain constraints on the function $f$, and would like to determine whether any function which satisfies those constraints and maps $\X$ to $\Y$. This leads to the key question in metric transforms: 
 \begin{question}For a given semi-metric space $\X$ and a given semi-metric space $\Y$, what is the full classification of functions $f$ that transform $\X$ to $\Y$?\end{question}
 
Much work has been done on metric transforms in the special case where ${\cal X}$ and ${\cal Y}$ are both Euclidean distances\footnote{When we refer to Euclidean or Manhattan distance in the remainder of this section, we always refer to distances in infinite dimensional Euclidean metric space and infinite dimensional Manhattan metric spaces, respectively.} or close variants.   
 Building on Schoenberg and Von Neumann's work~\cite{ns41}, Schoenberg~\cite{s38} classified all functions that transform Euclidean distances to Euclidean distances. 
 Interestingly, it is known that there is a close connection between these metric transforms and positive definite Euclidean kernels~\cite{s42, ss01}

 One natural question arises: what is the theory of metric transforms for non-Euclidean metrics? Surprisingly little attention has been paid to this question.
 
In the case when ${\cal X}$ is Manhattan (or $\ell_1$) distance, and ${\cal Y}$ is Euclidean distance, Schoenberg~\cite{s38} provided a partial categorization of functions that transform Manhattan distance to Euclidean distance. This was followed by Assouad's work in 1980, which provided a partial categorization of functions that transform Manhattan distances to Manhattan distances~\cite{a80}. Our work on metric transforms completes the partial categorizations of Schoenberg and Assouad, and proves their partial categorization is a full categorization.

\subsubsection{Main Result}

Our main result about metric transforms is a complete classification of functions that transform Manhattan distances to Manhattan distances. First, we need to define Bernstein functions: 
\begin{definition}[Bernstein functions~\cite{b29}]\label{def:bernstein}
  A function $f:\mathbb R_{\geq 0}\to \mathbb R_{\geq 0}$ is Bernstein if $f(0)=0$ and its derivative $f'$ is completely monotone (see Definition~\ref{def:cm}) when restricted to $\mathbb R^+$. 
  Equivalently, a function $f$ is Bernstein if:
\begin{compactenum}
\item $(-1)^k \frac{\d^k f(x)}{\d x^k} \leq 0$ for all $k \geq 1, x \geq 0$,
\item $f(x) \geq 0$ for all $x \geq 0$, and
\item $f(0) = 0$.\footnote{We remark that the special attention on $f(0)$ in the definitions above is a bit non-standard but are convenient for our purposes.}
\end{compactenum}
\end{definition}
Now we are ready to state our main result:

\begin{theorem}[Main result, classifying all Manhattan metric transforms, informal version and combination of Theorem~\ref{thm:formal_manhattan_transform_1_and_2} and \ref{thm:formal_manhattan_transform_2_and_3}] \label{thm:informal_manhattan_transform}

  For a function $f:\mathbb R_{\geq 0}\to\mathbb R_{\geq 0}$, the following are equivalent:

  \begin{compactenum}
    \item $f$ is Bernstein.
    \item $f$ transforms Manhattan distances to
  Manhattan distances.
    \item  $f$ transforms Manhattan distances to squared Euclidean distances.
  \end{compactenum}
\end{theorem}

It was previously known that Bernstein functions transform Manhattan distances to Manhattan distances~\cite{a80}, and that they transform Manhattan distances to squared Euclidean distances~\cite{s38}, but in both cases, it was not previously known that these were the only such functions. It was previously conceivable that, in situations where one needs a metric transform involving Manhattan spaces, but Bernstein functions do not suffice, one could find other suitable metric transforms; our Theorem~\ref{thm:informal_manhattan_transform} rules out such a possibility. This also has a number of simple consequences, for instance: given any $n$ points $x_1, \ldots x_n$ in the metric space ($\R^d, \ell_1$) for any $d$, one can use our construction in Theorem~\ref{thm:informal_manhattan_transform} to explicitly calculate $F: \R^d \to \ell_1$ such that $\|F(x_i)-F(x_j)\|_1 = f(\|x_i-x_j\|_1)$.

\subsection{Only Polynomials Preserve Low Rank Matrices} \label{sec:polymeth}

The \emph{polynomial method} is a powerful technique for designing algorithms and constructing combinatorial objects. A key insight behind many of these results is the following fact, that applying a low-degree polynomial entry-wise to a low-rank matrix yields another low-rank matrix:
\begin{fact}[The polynomial method, folklore; see e.g.~\cite{clp17}] \label{fact:polymethod}
  Suppose $f : \R \to \R$ is a polynomial of degree $d$. Then, for any matrix $M \in \R^{n \times n}$ of rank $r$, the matrix $M^f \in \R^{n \times n}$ given by $M^f_{i,j} := f(M_{i,j})$ has $\rank(M^f) \leq 2\binom{r+\lfloor d/2 \rfloor-1}{\lfloor d/2 \rfloor}$. For instance, if $r = \log_2 n$ and $d < o(\log_2 n)$, then $\rank(M^f) < n$.
\end{fact}

For one example, consider the fastest known algorithm for batch Hamming Nearest Neighbor Search due to Alman, Chan, and Williams~\cite{acw16}. In this problem, one is given as input $2n$ vectors $x_1, \ldots, x_n, y_1, \ldots, y_n \in \{0,1\}^d$ for $d = \Theta(\log n)$, and a threshold value $t \in \{0,1,\ldots,d\}$, and one wants to find a pair $(i,j) \in [n] \times [n]$ such that the Hamming distance between $x_i$ and $y_j$ is at most $t$. \cite{acw16} takes an algebraic approach to this problem, by first considering the matrix $M \in \R^{n \times n}$ where $M_{i,j}$ is the Hamming distance between $x_i$ and $y_j$. One can see that $\rank(M) \leq 2d$, and one could use fast matrix multiplication to quickly compute all the entries of $M$\footnote{\label{foot1}We first construct the matrices $X \in \R^{n \times 2d}$ and $Y \in \R^{2d \times n}$ such that $M = X \times Y$. We can then compute the product $X \times Y$ in $\tilde{O}(n^2)$ time using fast rectangular matrix multiplication~\cite{c82,w18} as long as $d < n^{0.1}$.}. However, since $M$ itself has $n^2$ entries, this could not improve much on the straightforward $O(n^2 \log n)$ time algorithm. They instead take the following approach:
\begin{enumerate}
    \item Pick a parameter $g = n^\delta$ for a constant $\delta>0$, and a function $f : \R \to \R$ such that $f(x) > g^2$ for all $x \in \{0,1,\ldots,t\}$, and $f(x) \in [0,1]$ for all $x \in \{t+1, t+2, \ldots, d\}$. \cite{acw16} use Chebyshev polynomials to construct such an $f$ which is a low-degree polynomial, so that the matrix $M^f$ has low rank by Fact~\ref{fact:polymethod}.
    \item Let $S_1, \ldots, S_{n/g}$ be a partition of $[n]$ into $n/g$ groups of size $g$, and consider the matrix $F \in \R^{\frac{n}{g} \times \frac{n}{g}}$ given by $F_{a,b} = \sum_{i \in S_a} \sum_{j \in S_b} M^f_{i,j}$. It is not hard to verify that $\rank(F) \leq \rank(M^f)$. Moreover, by the way $f$ was defined, an entry $F_{a,b}$ is larger than $g^2$ if and only if there is an $(i,j) \in S_a \times S_b$ such that the Hamming distance between $x_i$ and $y_j$ is at most $t$.
\end{enumerate}
There is a trade-off between the parameter $\delta$ and the degree of $f$, and hence the rank of $F$. \cite{acw16} balance this trade-off to yield a matrix $F$ of low rank\footnote{They pick rank $\approx n^{0.1}$ in order to apply fast rectangular matrix multiplication as in footnote~\ref{foot1}, although different applications of the polynomial method have aimed for different target ranks.} and dimensions $n^{1-\delta} \times n^{1-\delta}$ for some $\delta>0$. Since $F$ now has a subquadratic total number of entries, fast matrix multiplication can be used to compute all its entries and solve the problem, in roughly $O(n^{2-2\delta})$ time.

The polynomial method in algorithm design is used like this to design the fastest known algorithms for a variety of different, important problems, including: the Orthogonal Vectors problem from fine-grained complexity~\cite{awy14,cw16}, All-Pairs Shortest Paths~\cite{w18,cw16}, the lightbulb problem in which one wants to find a planted pair of correlated vectors among a collection of random vectors~\cite{v12,kkk18,a18}, computational problems related to kernel methods in spectral clustering and semi-supervised learning~\cite{acss20}, and some stable matching problems~\cite{mps16}. In all these works, one starts with a matrix $M$ describing the input data which has low rank, and one transforms it into a matrix like $M^f$ which `amplifies' the key properties of the data while still having low rank. A similar approach has also been used to bound the ranks of matrices which arise in other settings, such as in the recent resolution of the Cap Set Conjecture from extremal combinatorics~\cite{clp17, eg17}, and in recent proofs that Hadamard and Fourier transforms have low Matrix Rigidity~\cite{aw17, dvir2017matrix, dvir2019fourier}.

This motivates the question: 
\begin{question}\label{question:polys}
Is it possible to generalize the polynomial method (Fact~\ref{fact:polymethod}) to functions $f$ other than polynomials?
\end{question}
In other words, are there functions $f$ which are not polynomials, but such that if one starts with any low-rank matrix $M$, and applies it entry-wise yielding the matrix $M^f$, then $M^f$ also has low rank? This would allow algorithm designers to expand the efficacy and reach of the polynomial method, both by expanding the set of constraints on the function $f$ (such as those in step 1 of the algorithm above) that one could use in the recipe above, and by potentially allowing us to find new functions which satisfy those constraints but lead to lower rank bounds, and hence faster algorithms.

\paragraph{Application to Transformers in NLP} Question~\ref{question:polys} is also important in the study of \emph{transformers} in machine learning. Transformers are a type of neural network structure that has been widely applied to many natural language processing (NLP) tasks \cite{vsp+17}. A common computational task which arises when training transformers is to calculate the `self attention' \cite{transformers}; formally, in this task, we are given three matrices $ A, B , C \in \R^{n \times d}$ where $n \gg d$,\footnote{The matrices $A, B$ and $C$ correspond to the query, key, and value matrices, respectively, when training transformers in NLP applications. For more background, we refer the reader to the post by Kulshrestha~\cite{transformers} and more followups \cite{kkl19,cld+20,fzs21,wlk+20}.} and we would like to compute
\begin{align}\label{eq:transformer}
    (A  B^\top)^f \cdot C
\end{align}
where $f : \R \rightarrow \R$ is a non-linear function that we apply entry-wise to the matrix $A B^\top \in \R^{n \times n}$, then we multiply the result on the right by $C$. In many applications, $f$ is the soft-max function. 

Naively evaluating Eq.~\eqref{eq:transformer} takes time $O( n^2 d )$ (without using fast matrix multiplication).  
However, if we can quickly find matrices $\wt{A}, \wt{B} \in \R^{n \times \wt{d} }$ for some $\wt{d} < n$ such that
\begin{align*}
    (AB^\top)^f = \wt{A} \times \wt{B}^\top,
\end{align*}
then we can evaluate Eq.~\eqref{eq:transformer} more quickly by first computing $\wt{B}^\top \times C$ and then computing $\wt{A} \times (\wt{B}^\top \times C)$, for a total running time of just $O(n d \wt{d})$.

Since $AB^\top$ can be any rank $d$ matrix, and $\wt{A} \times \wt{B}^\top$ has rank at most $\wt{d}$, it follows that an upper bound on the best $\wt{d}$ we can achieve is the maximum, over all matrices $M$ of rank $d$, of $\rank(M^f)$. Question~\ref{question:polys} asks whether it is possible to achieve $d' < n$ for functions $f$ like the soft-max function which are not a polynomial. If not, then we can only hope to carry out this plan of attack if we can find a low-degree polynomial approximation to our function $f$.

\subsubsection{Main Result}

More formally, the functions $f$ we are interested in are those which preserve low-rank matrices. We first define applying a function to a matrix entry-wise, then the matrices of interest.
\begin{definition}[Entry-wise application]
For a function $f : \R \to \R$ and matrix $M \in \R^{a \times b}$, the \emph{entry-wise application of $f$ to $M$} is the matrix $M^f \in \R^{a \times b}$ where $M^f_{i,j} := f(M_{i,j})$, for $(i,j) \in [a] \times [b]$.
\end{definition}

\begin{definition}[Preserve low-rank matrices] \label{def:preservelowrank}
For a function $f : \R \to \R$ and positive integer $n$, we say $f$ \emph{preserves low-rank $n \times n$ matrices} if, for every matrix $M \in \R^{n \times n}$ with $\rank(M) \leq \lceil \log_2(n) \rceil + 1$, we have $\rank(M^f) < n$.
\end{definition}

For a function $f$ to be effective in the polynomial method as described above, it is necessary (but usually not sufficient) that $f$ preserves low-rank $n \times n$ matrices in the sense of Definition~\ref{def:preservelowrank}. Indeed, in all the aforementioned applications of the polynomial method, such as the algorithm of~\cite{acw16} and the application to transformers that we described above, the original matrix $M$ describing the data can have rank greater than $\log_2 n$. The details of how low the rank of $M^f$ must be can vary in the different applications, but it is always necessary that $M^f$ has less than \emph{full} rank (i.e., $\rank(M^f) < n$). 

Our main result answers Question~\ref{question:polys} in the negative, showing that Fact~\ref{fact:polymethod} cannot be generalized. 

\begin{theorem}[Main result, informal statement of Theorem~\ref{thm:log_formal}]\label{thm:log}
For any positive integer $n \geq 2$, if the real analytic function $f : \R \to \R$ preserves low-rank $n \times n$ matrices, then $f$ is a polynomial of degree at most $\lceil \log_2(n) \rceil$.
\end{theorem}

This shows that real analytic functions $f$ which are not polynomials do not preserve low-rank $n \times n$ matrices, and only polynomials of degree less than $\lceil \log_2(n) \rceil$ can preserve low-rank $n \times n$ matrices.
Hence, one cannot hope to improve on the polynomial method by extending it to any classes of real analytic functions other than low-degree polynomials. 

We note that there is a small constant-factor gap between the degree which Fact~\ref{fact:polymethod} tells us is sufficient for a polynomial to preserve low-rank $n \times n$ matrices, and the degree that Theorem~\ref{thm:log} says is necessary: for instance, Fact~\ref{fact:polymethod} says that polynomials of degree at most $\frac12 \log_2(n)$ suffice, since $\binom{\frac54 \log_2(n)}{\frac14 \log_2(n)} \ll n$, whereas Theorem~\ref{thm:log} says that degree less than $\log_2(n)$ is necessary. We leave open the question of closing this gap, although we note that the constant factor in front of the polynomial degree does not play a major role in most of the aforementioned applications of Fact~\ref{fact:polymethod}.\footnote{For instance, our running example algorithm of~\cite{acw16} only uses an asymptotic bound on how the degree grows with the dimension of the input points, and the constant factor in front of the polynomial degree is ultimately subsumed by a `$O$' in the running time.}

\section{Technique Overview} \label{sec:techniques}

\subsection{Key Lemma}
In this section, we introduce our key new technical idea, representation theory of the real hyperrectangle. This technique computes eigenvectors and eigenvalues of a large class of matrices which are defined in terms of hyperrectangles. We first describe and provide intuition about this technique, then we will explain how it leads to our applications by demonstrating why these matrices and their eigenvalues are relevant to kernel methods, metric transforms, and the converse of the polynomial method.

Our new technique concerns matrices defined in terms of a real hyperrectangle. 
\begin{definition} [Real hyperrectangle]\label{def:hyperrectangle}
The $d$-dimensional real hyperrectangle parameterized by $d$ variables $a_1, \ldots a_d > 0$ is the convex hull of the $2^d$ points $\{\pm a_1/2, \ldots \pm a_d/2 \}$.
\end{definition}

The eigenvectors of the family of matrices we define shortly will come from columns of Walsh-Hadamard matrices.
\begin{definition} [Walsh-Hadamard matrices]
For a positive integer $d$, let $v_1, \ldots v_{2^d} \in \{0,1\}^d$ be the enumeration of all $n$-bit vectors in lexicographical order. The \emph{Walsh-Hadamard} matrix $H_d$ is the $2^d \times 2^d$ matrix defined by $H_d(v_i, v_j) := (-1)^{\langle v_i, v_j \rangle }$, where $\langle v_i, v_j \rangle$ is the inner product between $v_i$ and $v_j$.
\end{definition}

We now introduce our key new technical lemma:

\begin{lemma} [Representation Theory of the Real Hyperrectangle, informal version of Lemma~\ref{lem:fourier_formal}]\label{lem:fourier_informal} 
Consider a $d$-dimensional hyperrectangle (Definition~\ref{def:hyperrectangle}) parameterized by $a_1, \ldots a_d > 0$. Enumerate the vertices in lexicographical ordering as $p_1, \ldots p_{2^d}$.
 
For any $f: \R \to \R$, let $D$ be the $2^d$ by $2^d$ matrix given by $D_{i,j} =f(\| p_i - p_j \|_1)$. Then:
 \begin{enumerate}
     \item $\Sigma := H_d D H_d$ is a diagonal matrix whose entries are the eigenvalues of $D$ multiplied by $2^d$, and $D = 4^{-d} \cdot H_d \Sigma H_d $.
     \item Let $v_1, \ldots v_{2^d}$ be the columns of the Hadamard matrix $H_d$. Then $v_1, \ldots v_{2^d}$ are the eigenvectors of $D$. For $i \in [2^d]$, let $B(i) \in \{0,1\}^d$ be the binary representation of $i$. Then, the eigenvalue corresponding to $v_i$ is: 
\begin{align}\label{eq:eigenvalue_formula}
\lambda_i = \sum_{b \in \{0,1\}^d} (-1)^{\langle B(i), b \rangle} \cdot f(\langle b, a \rangle). 
\end{align}
\end{enumerate}
\end{lemma}

We will see shortly that this expression for the eigenvalue $\lambda_i$ can also be rewritten in terms of integrals and derivatives of the function $f$, allowing us to use analytic techniques when computing or applying these eigenvalues.

\paragraph{Warm-up: $d$-dimension}  
To illustrate Lemma~\ref{lem:fourier_informal}, consider the case when the dimension of the hyperrectangle is $d=2$, and the hyperrectangle is parameterized by $a,b>0$. 

Let 
\begin{align*}
p_1 = (+a/2,+b/2), ~ 
p_2 = (-a/2,+b/2), ~ 
p_3 = (+a/2,-b/2), ~ 
p_4 = (-a/2,-b/2)
\end{align*}
be the vertices of the hyperrectangle. 

The matrix $D \in \R^{4 \times 4}$ we consider is defined by $D_{i,j} = f(\|p_i - p_j\|_1)$, and is thus given by:
\begin{align}\label{eq:matrix}
D = \begin{bmatrix} 
f(0) & f(a) & f(b) & f(a+b) \\
f(a) & f(0) & f(a+b) & f(b) \\
f(b) & f(a+b) & f(0) & f(a) \\
f(a+b) & f(b) & f(a) & f(0)
\end{bmatrix}.
\end{align}
Lemma~\ref{lem:fourier_informal} says that $D$'s eigenvectors are the columns of the $4$ by $4$ Hadamard matrix $H_2$:
\begin{align*}
v_1= \begin{bmatrix} +1 \\ +1 \\ +1 \\ +1 \end{bmatrix}, 
  v_2 = \begin{bmatrix} +1 \\ -1 \\ +1 \\ -1 \end{bmatrix}, 
  v_3 = \begin{bmatrix} +1 \\ +1 \\ -1 \\ -1 \end{bmatrix}, 
  v_4 = \begin{bmatrix} +1 \\ -1 \\ -1 \\ +1 \end{bmatrix}.
\end{align*}

We can verify that these are the eigenvectors, and compute the corresponding eigenvalues 
$\lambda_1, \lambda_2, \lambda_3, \lambda_4$, by multiplying the first row of $D$ by
$v_1, v_2, v_3, v_4$:
\begin{align}\label{eq:eigval}
\nonumber&\lambda_1 = f(0) + f(a) + f(b) + f(a+b),\\
\nonumber&\lambda_2 = f(0) - f(a) + f(b) - f(a+b),\\
&\lambda_3 = f(0) +f(a) - f(b) - f(a+b),\\
\nonumber & \lambda_4 = f(0) - f(a) - f(b) + f(a+b).
\end{align}
This is the $d=2$ version of Eq.~\eqref{eq:eigenvalue_formula}.

A key remark in some of our proofs is that, if $f$ is smooth, then $\lambda_2, \lambda_3, \lambda_4$ can also be written in terms of integrals using the fundamental theorem of calculus:
\begin{align}\label{eq:int}
\nonumber & \lambda_2 = -\int_0^a f'(x) dx - \int_b^{a+b} f'(x) \d x, \\
 & \lambda_3 = -\int_0^b f'(x) \d x - \int_a^{a+b} f'(x) \d x, \\
\nonumber & \lambda_4 = \int_0^a \int_0^b f''(x+y) \d x \d y. 
\end{align}
Expressions similar to Eq.~(\ref{eq:int}) hold for the general, $d$-dimensional setting as well.

\paragraph{Proof idea: Representation Theory of the Real Hyperrectangle} We call our technique `representation theory of the real hyperrectangle' since it is proved by using representation theory to calculate the eigenvalues of a large class of matrices. Representation theory in general is used to calculate eigenvalues of matrices associated with objects that have group symmetry~\cite{fh91, etingof}. The $d$-dimensional real hyperrectangle has reflection symmetry about each of its $d$ axes, and Lemma~\ref{lem:fourier_informal} follows from analyzing this symmetry using Schur's Lemma from representation theory. In other words, Lemma~\ref{lem:fourier_informal} can be seen as a use of representation theory of the symmetry group of the real hyperrectangle. For more details on representation theory, see Lemma~\ref{lem:known-abelian} in Appendix~\ref{sec:preli:representation}. For a proof of Lemma~\ref{lem:fourier_informal}, see Appendix~\ref{sec:key}.

\paragraph{Related Work}
Representation Theory is a mathematical field dating back a hundred years, with many applications in physics and computer science. Representation theory is used in physics to calculate the spectra of Hamiltonians and compute molecular and atomic orbitals~\cite{feynman}. 

In computer science, representation theory is used to compute the vibrational spectra of graph Laplacians where the underlying graph has vertex-transitive group symmetry, a case which covers the boolean cube, cycle, buckyball, and other molecular structures~\cite{GS92, ODonnell14, spielman-notes}. Guattery and Miller implicitly used representation theory to give structure to the spectra of graphs where there exists a vertex automorphism of order two ~\cite{GM98}.  Representation theory is also central to the study of quantum tomography~\cite{OW16}, Boolean function analysis~\cite{ODonnell14}, low-sparsity expander construction~\cite{M88, LPS88}, random walk theory~\cite{Diaconis02, Saloff04, FOW18}, and more. 

Representation Theory is closely related to Fourier transforms \cite{t10}, which have been extensively studied in theoretical computer science \cite{hikp12a,hikp12b,ikp14,ik14,m15,ps15,ckps16,k16,k17,nsw19,jls20,cm21}.

\paragraph{Use in Applications}

We next give an overview of how we use Lemma~\ref{lem:fourier_informal} to derive our three applications. We focus on explaining how the matrices described by Lemma~\ref{lem:fourier_informal} and their eigenvalues arise in each setting. At a high level, the class of matrices described by Lemma~\ref{lem:fourier_informal} is sufficiently general that we are able to show it is `complete' for the matrices or distance functions arising in our applications. At the same time, Lemma~\ref{lem:fourier_informal} allows us to easily compute the eigenvalues of these matrices. To our knowledge, a general enough class of matrices which capture our applications but whose eigenvalues are understood was previously not known, and this is what allows us to prove that previous partial categorizations (of positive definite kernels (Definition~\ref{def:euclineankernel}, \ref{def:manhattan_kernel}), metric transforms (Definition~\ref{def:metric-transform}), and functions which preserve low-rank (Definition~\ref{def:preservelowrank})) are in fact complete classifications.

\subsection{Kernel Methods}\label{sec:tech_kernel}

We begin with an overview of our proof of Theorem~\ref{thm:informal_kernel_manhattan}. Given any $n$ points in $d$-dimensional Manhattan space, it is known they can be isometrically embedded into $\ell_1$ restricted to the corners of some (possibly high dimensional) hyperrectangle~\cite{dl09}. Therefore, to prove Theorem~\ref{thm:informal_kernel_manhattan}, it suffices to find all functions $f$ such that the matrix $M \in \R^{2^d} \times \R^{2^d}$ defined as:
 \begin{align*}
  M_{i,j} = f(\|p_i - p_j\|_1) 
 \end{align*}
  is positive semi-definite whenever $p_1, \ldots p_{2^d}$ are the vertices of some hyperrectangle.

Fortunately, Lemma~\ref{lem:fourier_informal} gives us a closed form expression for the eigenvalues of $M$. For $M$ to be positive semi-definite, the eigenvalues of $M$ must all be nonnegative. We exploit a connection between eigenvalues of $M$ (which are computed by Eq.~(\ref{eq:eigenvalue_formula})) and discrete derivatives of $f$ to prove that $f$ must be completely monotone (Definition~\ref{def:cm}). The details are quite technical; for more details, see Appendix~\ref{sec:bern}. 
\subsection{Metric Transforms}

We next sketch the proof of Theorem~\ref{thm:informal_manhattan_transform}. Schoenberg~\cite{s38} previously showed that Bernstein functions transform Manhattan distances to squared Euclidean distances, and Assouad~\cite{a80} previously showed that Bernstein functions transform Manhattan distances to Manhattan distances. It thus suffices to prove that any function that transforms Manhattan to squared Euclidean must be Bernstein, and similarly for any function that transforms Manhattan to Manhattan.

\paragraph{Bernstein functions transform Manhattan to squared Euclidean}
Our starting point is a classical criterion for determining whether a set of distances is a squared Euclidean distance due to Schoenberg~\cite{s35}:

\begin{lemma} [Squared Euclidean distance criterion~\cite{s35}]
Given a set of distances $d_{i,j}$ for all $(i,j) \in [n] \times [n]$ satisfying $d_{i,j} = d_{j,i}$ and $d_{i,i} = 0$, then $d_{i,j}$ can be embedded into squared Euclidean distance if and only if matrix $D$ with $D_{i,j} = d_{i,j}$ satisfies $x^{\top} D x \leq 0$ for all $x \bot 1$.
\end{lemma}
This criterion $D$ must satisfy is known as the \emph{negative type condition}~\cite{dl09}. As in Section~\ref{sec:tech_kernel} above, we also know that any Manhattan distance can be isometrically embedded into Manhattan distances between a subset of the corners of a (possibly high) dimensional real hyperrectangle. Therefore, by carefully considering the definition of Bernstein functions, one can show: to prove that only Bernstein functions transform Manhattan to squared Euclidean, it suffices to show that the matrix $D$ where
 \begin{align*}
  D_{i,j} = f(\|p_i - p_j\|_1) 
 \end{align*}
 satisfies $x^{\top} D x \leq 0 $ for all $x \bot 1$, whenever $p_1, \ldots p_{2^d}$ are vertices of some hyperrectangle. 
 
 Lemma~\ref{lem:fourier_informal} tells us that that the all ones vector $v_1$ is an eigenvector of $D$, since $v_1$ is the first column of the Hadamard matrix. Therefore, it suffices to show that the eigenvalues of $D$ except for $\lambda_1$ are negative. We once again exploit a connection between eigenvalues of $D$ and discrete derivatives of $f$ to prove that $f$ must be Bernstein (Definition~\ref{def:bernstein}). For more details, see  Theorem~\ref{thm:formal_manhattan_transform_1_and_2} in Appendix~\ref{sec:manhattan_transform12}.

\paragraph{Manhattan to Squared Euclidean $\Leftrightarrow$ Manhattan to Manhattan}
We next show that functions that transform Manhattan to squared Euclidean must transform Manhattan to Manhattan, and vice versa. It is known that Manhattan distances isometrically embed into squared Euclidean distances~\cite{s38, dl09}, which implies that functions that transform Manhattan to Manhattan must transform Manhattan to squared Euclidean. 

To prove the other direction, suppose function $f$ transforms Manhattan to squared Euclidean. Consider as before the $2^d \times 2^d$ matrix $D$ where $D_{i, j} = f(\|p_i - p_j\|_1)$, where $p_1, \ldots p_{2^d}$ are vertices in lexicographical order of some real hyperrectangle (Definition~\ref{def:hyperrectangle}). 

Using the fact that $D$ contains squared Euclidean distances, we can explicitly find $2^d$ points whose pairwise squared Euclidean distances are the entries in $D$ (by combining Lemma~\ref{lem:fourier_informal} and methods of Schoenberg~\cite{s35}). We show that these $2^d$ points, themselves, lie on another $2^d$-dimensional real hyperrectangle! One can see using the Pythagorean theorem that squared Euclidean distances on the real hyperrectangle can be realized as Manhattan distances, so this shows that $f$ transforms Manhattan to Manhattan as well. For more details, see Appendix~\ref{sec:manhattan_transform23}.

\subsection{Polynomial Method Converse}
In this section, we sketch our techniques for Theorem~\ref{thm:log}. We also explain how the matrices from hyperrectangles in Lemma~\ref{lem:fourier_informal} arise in our methods.

Let $d = \lceil \log_2 n \rceil $. Suppose $M: \R^d \to \R^{2^d \times 2^d}$ is a family of matrices defined by $M(a) = \|p_i - p_j\|_1$ where $p_1, \ldots p_{2^d}$ are vertices in lexicographical order of the hyperrectangle paramaterized by $a_1, \ldots a_d$. We show that these matrices have rank at most $d+1$. Therefore if $f$ preserves low rank $n \times n$ matrices, then $M(a)^f$ must have rank $< n$ for all $a \in \R^d$.   

Recall that representation theory of the real hyperrectangle (Lemma~\ref{lem:fourier_informal}) gives an algebraic formula for the eigenvalues $\lambda_1^f(a), \ldots \lambda_{2^d}^f(a)$ of $M(a)^f$ in terms of $a$ and $f$. The fact that $M(a)^f$ does not have full rank for any $a$ means that, for every $a$, there is an $i$ such that $\lambda_{i}^f(a) = 0$. However, using Lemma~\ref{lem:fourier_informal}, we prove the stronger statement that there exists an $i$ such that for all $a$, we have $\lambda_{i}^f(a) = 0$.

Next, we show that if $\lambda_{i}^f(a) = 0$ for all $a$, then $f^{(d)}(x)=0$ for all $x \in \R$, where $f^{(d)}$ represents the $d^{th}$ derivative of $f$. We do this by writing $f^{(d)}$ as a linear combination of $\lambda^f_{i}(a)$ for various settings of $a$, making use of our integral expressions similar to Eq.~(\ref{eq:int}) above. This implies that $f$ is a degree $d=\lceil \log_2 n \rceil$ polynomial if it preserves low rank.

\newpage
\tableofcontents 
\newpage
\bibliographystyle{alpha}
\bibliography{ref.bib}

\newpage
\appendix
\section*{Appendix}

\paragraph{Roadmap.}
In Section~\ref{sec:preli}, we define notations, provide several basic definitions and fundamental tools. In Section~\ref{sec:non_polynoimial}, we prove that non-polynomial function blows up the matrix rank. It proves Theorem~\ref{thm:informal_kernel_manhattan}. In Section~\ref{sec:manhattan_transform12}, we prove condition 1 and condition 2 in Theorem~\ref{thm:informal_manhattan_transform} are equivalent. In Section~\ref{sec:manhattan_transform23}, we prove condition 2 and condition 3 in Theorem~\ref{thm:informal_manhattan_transform} are equivalent. Overall, Section~\ref{sec:manhattan_transform12} and Section~\ref{sec:manhattan_transform23} together prove Theorem~\ref{thm:informal_manhattan_transform}. In Section~\ref{sec:bern}, we have a proof of Theorem~\ref{thm:informal_kernel_manhattan}.
In Section~\ref{sec:key}, we prove our result about representation theory of real hyperrectangles.

\section{Preliminaries}\label{sec:preli}

This section is organized as follows:

\begin{itemize}
    \item In Section~\ref{sec:preli:notation}, we define several basic notations.
    \item In Section~\ref{sec:preli:definition}, we provide some definitions about Hadamard matrix, high-dimensional hyperrectangle.
    \item In Section~\ref{sec:preli:classification}, we provide some previous work about the classifications of completely monotone and Bernstein function.
    \item In Section~\ref{sec:preli:metric}, we state well-known results about metric hierarchies.
    \item In Section~\ref{sec:preli:negative}, we define negative metrics and euclidean embeddability.
    \item In Section~\ref{sec:preli:representation}, we present previous work about representation theory tools.
\end{itemize}

\subsection{Notations}\label{sec:preli:notation}
For a vector $x$, we use $\| x \|_1$ to denote the entry-wise $\ell_1$ norm of $x$. We use $\| x \|_2$ to denote the entry-wise $\ell_2$ norm of $x$. For two vectors $a,b$, we use $\langle a, b \rangle$ to denote the inner product between $a$ and $b$. For a vector $x$, we use $x^\top$ to denote the transpose of $x$.

\subsection{Definitions}\label{sec:preli:definition}
We provide an alternate but equivalent definition of $H_d$ as the square Hadamard matrix with $2^d$ rows. These matrices consist of $\pm 1$-valued entries and are defined recursively via:

\begin{align*}
H_0&=[1]\\
H_{k+1}&=\begin{bmatrix}H_k&H_k\\ H_k& -H_k\end{bmatrix}, \quad k\geq 0.
\end{align*}
 For a review of Hadamard matrices, see~\cite{hadamard}.

\paragraph{Hyperrectangles} Often times in our proof, we may say things like ``let $x_1, \ldots
x_{2^d}$ be the corners of a $d$ dimensional hyperrectangle''. For these statements
to make sense, we must specify which corner $x_i$ refers
to. Scale the $d$ dimensional hyperrectangle to be an axis-aligned hypercube, and place
one of the hypercube corners at the origin. Each corner then has a binary number $b$ as its coordinate bit string. We let $x_{b+1}$ refer to the original hyperrectangle
corner corresponding to $b$.

\subsection{Alternate Classifications of Completely Monotone and Bernstein Functions}\label{sec:preli:classification}

Here we recall the classical Bernstein Theorem from analysis constructively classifying completely monotone (Definition~\ref{def:cm}) and Bernstein functions (Definition~\ref{def:bernstein}).
\begin{prop}[Chapter 14, Theorems 3 and 6 in \cite{lax}]\label{prop:cm}

For a function $f:\mathbb R_{> 0}\to\mathbb R_{\geq 0}$, the following are equivalent:

\begin{enumerate}
  \item $f$ is completely monotone.
  \item Letting $(D_af)(x)=f(x+a)-f(x)$, for any $(a_1,\dots,a_n)$ non-negative we have 

  \[(-1)^n \left(\prod_{i=1}^n D_{a_i}\right)f(x)\geq 0\]

  for all $x>0$.
  \item There exists a positive finite measure $\mu$ on $\mathbb R_{\geq 0}$ such that 

  \[f(x)=\int_0^{\infty} e^{-tx}\d \mu(t),\quad x>0.\]
\end{enumerate}

\end{prop}

The part 2 of Proposition~\ref{prop:cm} is essentially the definition we gave for completely monotone, except that it does not assume any smoothness or even continuity a priori. The third shows that all completely monotone functions are in fact mixtures of decaying exponentials. From the above one easily derives a corresponding classification of Bernstein functions. If $f$ also has $0$ in its domain, then the above result applies the same way, however (with the same measure $\mu$ as in part $3$ of Proposition~\ref{prop:cm}) we have 
\begin{align*}
f(0)\geq \mu(\mathbb R_{\geq 0})
\end{align*}

since we did not require any continuity at $0$. 

\begin{prop}[Theorem 6.7 in \cite{harmonic}]
\label{prop:bern}
For a function $f:\mathbb R_{\geq 0}\to\mathbb R_{\geq 0}$ with $f(0)=0$, the following are equivalent:

\begin{enumerate}
  \item $f$ is Bernstein.
  \item Letting $(D_af)(x)=f(x+a)-f(x)$, for any $(a_1,\dots,a_n)$ non-negative we have 

  \[(-1)^n \left(\prod_{i=1}^n D_{a_i}\right)f(x)\leq 0,\quad x>0.\]

  \item There exists a positive measure $\mu$ on $\mathbb R^+$ and $a,b\geq 0$ such that 

  \[f(x)=a+bx+\int_{\mathbb R^+} (1-e^{-tx}) \d \mu(t),\quad x>0.\]

  Here $\mu$ must satisfy $\int_{\mathbb{R}_+} \min\{1,t\} \d \mu(t)<\infty.$
\end{enumerate}

\end{prop}

Due to the second criterion just above, Bernstein functions are also sometimes called \emph{completely alternating}. We remark that these results apply more generally in the setting of abelian semigroups, where the integral is taken over a measure on the space of positive characters. This general point of view is explained in~\cite[Chapter 6]{harmonic}, and applies, for instance, to the semigroup of compact subsets of $\mathbb R$ under union.

\subsection{Metric Hierarchies}\label{sec:preli:metric}
Here are well-known facts we will use throughout our proof:

    \begin{lemma}\label{lem:l1-hyperrectangle} For any $n$ points $x_1, \ldots x_n$ in $\ell_1$,
      there exist $n$ points $y_1, \ldots y_n$ such that $\|x_i-
      x_j\|_1 = \|y_i - y_j\|_1$, and $y_1, \ldots y_n$ are a subset of corners of a
      $d$ dimensional hyperrectangle for some $d$. 
    \end{lemma}
    \begin{proof} 
      This follows from the equivalence of the cut cone and $\ell_1$
      distance (Theorem 4.2.2 in~\cite{dl09}).
    \end{proof}
    \begin{lemma}\label{lem:l2-hyperrectangle}
The squared Euclidean distance between points in the corners of a hyperrectangle
    isometrically embeds into Manhattan distance.
    \end{lemma}
    \begin{proof} 
      This follows from the Pythagorean theorem.
    \end{proof}

\begin{lemma} \label{lem:l1-iso}
  Manhattan distances embed isometrically into squared Euclidean
  distances.
\end{lemma}
\begin{proof} This follows from Corollary 6.1.4 and Lemma 6.1.7
in~\cite{dl09}. \end{proof}

\subsection{Negative Type Metrics and Euclidean Embeddability}\label{sec:preli:negative}
We now present a criterion by Schoenberg~\cite{s35} on when
a metric is isometrically embeddable into squared Euclidean
distances\footnote{
We note that Schoenberg's criteria has a beautiful proof, which one can
find one direction of in \cite{note}.
}.

\begin{definition}[negative type] A matrix $D$ is  iff $x^{\top} D x
  \leq 0$ for all $x \bot 1$.
\end{definition}

\begin{lemma}[Schoenberg~\cite{s35}]\label{lem:euc}  
Consider $x_1,\ldots, x_n$ where $d_{i,j}$ is the distance between $x_i$ and $x_j$.  Let $D$ be an $n$ by $n$
  matrix where $D_{i,j} = d_{i,j}^2$.  The distances $d_{i,j}$ are
  isometrically embeddable into Euclidean space iff the matrix $D$ is
  negative type.
\end{lemma}

We note that if $D$ happens to have the all ones vector $\textbf{1}$ as an eigenvector, we have a simpler criterion for testing if $D$ is negative type:

\begin{lemma}[Schoenberg Variant]\label{lem:euc-variant}  
Consider $x_1,\ldots, x_n$ where $d_{i,j}$ is the distance between $x_i$ and $x_j$.  Let $D$ be an $n$ by $n$
  matrix where $D_{i,j} = d_{i,j}^2$.  
  
  If the all ones vector is an eigenvector of $D$, then the $d_{i,j}$ are
  isometrically embeddable into Euclidean space iff every eigenvalue of $D$, excluding the eigenvalue correseponding to the all ones vector, is non-positive.
\end{lemma}
\begin{proof}
Lemma~\ref{lem:euc-variant} follows from Lemma~\ref{lem:euc} and the fact that every symmetric matrix has an orthonormal set of eigenvectors.
\end{proof}
If $d_{ij}$ is isometrically embeddable into Euclidean space, we can find an
explicit embedding:

\begin{lemma}\label{lem:emb} Consider $x_1, \ldots
  x_n$ where $d_{i,j}$ is the distance between $x_i$ and $x_j$.  Let $D$ be the matrix where $D_{i,j} =
  d^2_{i,j}$. Let $\Pi$ be the projection matrix off the all ones
  vector, i.e., $\Pi$ can be expressed explicitly as $I-J/n$, where $J$ is the $n \times n$ all-ones matrix, and $I$ is
  identity matrix. 
  
  Let $M:=-\frac{1}{2}\Pi D \Pi$.
  
  If $y_1, \ldots y_n$ are such that $\|y_i - y_j\|_2 = d_{i,j}$ and
  $\sum_{i=1}^n y_i = 0$, then $M_{i,j} = \langle y_i, y_j \rangle$.
  Moreover, if $M = U^{\top}U$ for some $U$, then the
  columns of $U$ are an embedding of $x_1, \ldots x_n$ into Euclidean
  space.
\end{lemma}
This follows from Eq.~2 in~\cite{critchley}.  A longer exposition of the link
between distance matrices and inner product matrices can be found
in~\cite{critchley}.

\subsection{Useful Tools}\label{sec:preli:representation}

We present Schur's lemma for Abelian groups $G$. Schur's lemma is one of
the cornerstones of representation theory~\cite{etingof}.

\begin{lemma}[Schur's lemma for Abelian groups]\label{lem:known-abelian}  If $G$ is a finite Abelian group of $n \times n$ matrices
  under multiplication, and $M$ is an $n \times n$ diagonalizable matrix
  satisfying $Mg = gM$, for all $g \in G$, then there exists a set of linearly independent vectors $v_1, \ldots v_n$ that are eigenvectors of $M$ and all $g \in G$. In other words, $M$ and $G$ are simultaneously diagonalizable.
\end{lemma}
Schur's Lemma will be useful in proving our key result about representation theory of the real hyperrectangle, or Lemma~\ref{lem:fourier_informal}.

\section{Non-Polynomial Functions Blow Up Matrix Rank}\label{sec:non_polynoimial}

The major goal of this section is to prove Theorem~\ref{thm:log_formal}. This section is organized as follows
\begin{itemize}
    \item In Section~\ref{sec:poly_preli}, we show some basic facts.
    \item In Section~\ref{sec:poly_imply}, we show that one eigenvalue is identically zero.
    \item In Section~\ref{sec:zero_eig}, we prove that only polynomials have a zero eigenvalue.
    \item In Section~\ref{sec:eigsum}, we rewrite the sum of eigenvalues.
    \item In Section~\ref{sec:converge}, we show the convergence via calculating the limit.
    \item In Section~\ref{sec:poly_main}, we state and prove our main result.
\end{itemize}

\subsection{Preliminaries}\label{sec:poly_preli}

We start with defining a useful tool.
\begin{lemma}\label{lem:zeroset}
If $g_1, \ldots g_n: \mathbb{R}^d \rightarrow \mathbb{R}$ are all Taylor expandable, and the union of the zero-sets of $g_i$ is all of $\mathbb{R}^d$, then one of $g_1, \ldots g_n$ is identically zero.
\end{lemma}
\begin{proof}

Firstly, if $g:\mathbb{R}^d \rightarrow \mathbb{R}$ is Taylor expandable, then the zero-set of $g$ has a well-defined measure.  

Secondly, if the measure of the zero-set is non-zero, there must exist an open ball in which $g$ is $0$. If this is the case, every higher order derivative at the center of the ball must be $0$, meaning the Taylor series for that function is identically zero.

Finally, since the union of the zero-sets of $g_i$ is the entire plane, one of their zero-sets has non-zero measure. Thus,  
it must be identically zero. 
\end{proof}

\subsection{One Eigenvalue is Identically Zero}\label{sec:poly_imply}
The goal of this section is prove Lemma~\ref{lem:poly_imply}.
\begin{lemma}[One eigenvalue is identically zero] \label{lem:poly_imply}
For any Taylor expandable function $f$, any $n$, and $d:= \log n+1$: we can find $M:\mathbb{R}^d \rightarrow \mathbb{R}^{n \times n}$ and Taylor expandable $\lambda^f_i:\mathbb{R}^d \rightarrow \mathbb{R}$ satisfying:
\begin{enumerate} 
\item $M(a)$ has rank $\leq d$ for all $a \in \mathbb{R}^d$
\item $\lambda^f_1(a) \ldots \lambda^f_n(a)$ is the full set of eigenvalues of $f(M(a))$, for all $a \in \R^d$.
\item If there exists $i \in [n]$ such that $\lambda^f_i(a) = 0$ for all $a \in \mathbb{R}^d$, then $f$ is a degree $d \leq \log n + 1$ polynomial. 
\end{enumerate}
\end{lemma}

\begin{proof} 

{\bf Constructing $M$.}
Consider a mapping $B : \{0, 1, \ldots 2^{d}-1\} \rightarrow \{0,1\}^d$ corresponding to the conversion of integers into $d$-digit binary strings, which we interpret as $d$ dimensional $0-1$ vectors. We set
\begin{align*}
M(a)_{i,j} = \langle a, B(|i-j|) \rangle
\end{align*}
where $a \in \R^d$. 

{\bf Constructing $\lambda^f_i$.}
For each matrix $M(a) \in \R^{n \times n}$, we established previously that $f(M(a)) \in \R^{n \times n}$ has eigenvectors equal to the Hadamard matrix columns, and the corresponding eigenvalues are:  
\begin{align*}
\lambda^f_i(a) = \sum_{b \in \{0,1\}^d} (-1)^{\langle B(i), b \rangle} \cdot f(\langle b, a \rangle) 
\end{align*}
We note that if $f$ is Taylor expandable, then so is $\lambda_i^f$ for all $i$ and $f$.

As noted before, $\lambda^f_i$ is Taylor expandable if $f$ is Taylor expandable. Also, $\lambda^f_i(a)$ forms the full set of eigenvalues for $M(a)$. Therefore, all that's left to prove is that if any $\lambda^f_i$ is $0$, then so is $f^{(d)}$.

If $\lambda^f_i$ is $0$, then 
\[
 (-1)^{\langle B(i), 1 \rangle} \sum_{b \in \{0,1\}^d} (-1)^{\|b\|_1} \cdot \lambda^f_i(  a+\epsilon b  ) = 0
\]
since it is the sum and difference of $\lambda^f_i$ evaluated at various points. Now:
\begin{align*}
& ~ (-1)^{\langle B(i), 1 \rangle} \sum_{b \in \{0,1\}^d} (-1)^{\|b\|_1} \cdot \lambda^f_i(  a+\epsilon b  ) \\
= & ~ 
(-1)^{\langle B(i), 1 \rangle} \sum_{b_1 \in \{0,1\}^d} (-1)^{\|b_1\|_1} \cdot \left(\sum_{b_2 \in \{0, 1\}^d} (-1)^{\langle B(i), b_2 \rangle} f(\langle b_2, a + \epsilon b_1 \rangle) \right)\\
= & ~ (-1)^{2 \langle B(i), 1 \rangle} \sum_{b \in \{0, 1\}^d} (-1)^{\|b\|_1} \cdot f ( \langle a + \epsilon b, {\bf 1} \rangle)\\
= & ~ \sum_{b \in \{0, 1\}^d} (-1)^{\|b\|_1} \cdot f ( \langle a + \epsilon b, {\bf 1} \rangle
\end{align*}
where the first equality follows from the definition of $\lambda_i^f$ and the second equality follows from Lemma~\ref{lem:eigsum}. It follows that if $\lambda_i^f = 0$, then
\[
\sum_{b \in \{0, 1\}^d} (-1)^{\|b\|_1} \cdot f ( \langle a + \epsilon b, {\bf 1} \rangle = 0
\] 
for all $\epsilon$ and $a$. By taking the limit as $\epsilon \to 0$ and dividing by $\epsilon^d$, we have for all $a$:
\begin{align}\label{eq:final}
\lim_{\epsilon \rightarrow 0} \frac{1}{ \epsilon^d}\sum_{b \in \{0, 1\}^d} (-1)^{\|b\|_1} \cdot f ( \langle a + \epsilon b, {\bf 1} \rangle) = 0
\end{align}
By Lemma~\ref{lem:converge}, the LHS of Eq.~\eqref{eq:final} is $f^{(d)}(a)$, so $f^{(d)}(a) = 0$ for all $a$. Therefore, $f$ is at most a degree $d$ polynomial as desired. Thus, we complete the proof.
\end{proof}

\subsection{Only Polynomials Have a Zero Eigenvalue}\label{sec:zero_eig}

The goal of this section is to prove Lemma~\ref{lem:zero_eig}.
\begin{lemma}[Only polynomials have a zero eigenvalue]\label{lem:zero_eig}
Given:
\begin{enumerate} 
\item A function $f : \R \rightarrow \R$ 
\item A function $M: \mathbb{R}^d \rightarrow \mathbb{R}^{n \times n}$, mapping $d$ dimensional vectors to $n$ dimensional matrices.
\item A set of $n$ functions $\lambda^f_1, \lambda^f_2, \cdots, \lambda^f_n$
such that each $\lambda^f_i : \R^d \rightarrow \R$ is Taylor expandable, and $\lambda^f_1(a) \ldots \lambda^f_n(a)$ is the full set of eigenvalues of $f$ applied entry-wise to $M(a)$ for all $a \in \mathbb{R}^d$,
\end{enumerate}

Then if $f$ transforms matrices $M(a)$ to rank $< n$ for all $a \in \mathbb{R}^d$, then there exists $i \in [n]$ where function $\lambda^f_i = 0$.
\end{lemma}

\begin{proof} 
If $f$ transforms matrix $M(a)$ to rank $<n$, then for any $a$,  there exists an $i$ where $\lambda^f_i(a)=0$. Thus the union (over $i$) of the zero sets of $\lambda^f_i$ is $\mathbb{R}^d$. We can then apply Lemma~\ref{lem:zeroset} to show that one of $\lambda^f_i$ is identically $0$ as desired.
\end{proof}

\subsection{Rewriting the Sum}\label{sec:eigsum}

The goal of this section is to prove \ref{lem:eigsum}. 

\begin{lemma}[Rewriting the sum]\label{lem:eigsum}
\begin{align*}
& ~ \sum_{b_1 \in \{0,1\}^d} (-1)^{\|b_1\|_1} \left(\sum_{b_2 \in \{0, 1\}^d} (-1)^{\langle B(i), b_2 \rangle} f(\langle b_2, a + \epsilon b_1 \rangle) \right)\\
= & ~ (-1)^{\langle B(i), {\bf 1} \rangle} \sum_{b \in \{0, 1\}^d} (-1)^{\|b\|_1} \cdot f ( \langle a + \epsilon b, {\bf 1} \rangle  )
\end{align*}
where $a$ and $b$ are $d$-dimensional vectors.
\end{lemma}
\begin{proof}

First, we can show: 
If $b_2$ is a $d$ dimensional vector with any $0$s in its vector notation, we know
\begin{align}\label{eq:sum0}
\sum_{b_1 \in \{0,1\}^d} (-1)^{\|b_1\|_1} f(\langle b_2, a + \epsilon b_1 \rangle )  = 0
\end{align}
for any $\epsilon$, and any constant $d$ dimensional vector $a$. The reason is if $b_2$ has any $0$'s in its vector notation, then flipping the corresponding bit in $b_1$ causes $(-1)^{\|b_1\|_1}$ to change sign, while leaving $\langle b_2, a+\epsilon b_1 \rangle$ unchanged.

Now, we know that: 
\begin{align*}
& ~ \sum_{b_1 \in \{0,1\}^d} (-1)^{\|b_1\|_1} \left(\sum_{b_2 \in \{0, 1\}^d} (-1)^{\langle B(i), b_2 \rangle} f(\langle b_2, a + \epsilon b_1 \rangle) \right) \notag\\
= & ~ \nonumber
\sum_{b_2 \in \{0, 1\}^d} (-1)^{\langle B(i), b_2 \rangle}
 \left(\sum_{b_1 \in \{0,1\}^d} (-1)^{\|b_1\|_1} f(\langle b_2, a + \epsilon b_1 \rangle) \right)\\
= & ~ (-1)^{\langle B(i), {\bf 1} \rangle}
 \left(\sum_{b_1 \in \{0,1\}^d} (-1)^{\|b_1\|_1} f(\langle {\bf 1}, a + \epsilon b_1 \rangle) \right).
\end{align*}
where the first equality follows by rearranging sums, and the second equality follows from Eq.~\eqref{eq:sum0}. This completes the proof. 

\end{proof}

\subsection{Calculating the Limit}\label{sec:converge}

The goal of this section is to prove Lemma~\ref{lem:converge}.
\begin{lemma}[Calculating the limit]\label{lem:converge}

Suppose the $d^{th}$ derivative of $f$, denoted as $f^{(d)}$, is continuous. Then:
\begin{align*}
\lim_{\epsilon \rightarrow 0} \epsilon^{-d}  \sum_{b \in \{0, 1\}^d} (-1)^{\|b\|_1} \cdot f ( \langle a + \epsilon b, {\bf 1} \rangle  ) = f^{(d)}( \langle a , {\bf 1} \rangle ).
\end{align*}
\end{lemma}
\begin{proof}

We have:
\begin{align} \label{eq:simplified}
\sum_{b \in \{0, 1\}^d} (-1)^{\|b\|_1} \cdot f ( \langle a + \epsilon b, {\bf 1} \rangle  ) 
= & ~  \sum_{s=0}^d (-1)^s \binom{d}{s} \cdot f ( \langle a + \epsilon b, {\bf 1} \rangle  ) \notag \\
= & ~  \sum_{s=0}^d (-1)^s \binom{d}{s} \cdot f( \langle a , {\bf 1} \rangle + s \epsilon )  \notag \\
= & ~ \int_0^\epsilon \int_0^\epsilon \ldots \int_0^\epsilon f^{(d)}( \langle a+ x, {\bf 1}\rangle ) \d x_1 \ldots \d x_d 
\end{align}
which we note, is independent of $i$.  The first and second equality follow from grouping $b$ by the number of ones it has, which we denote as $s$. The last equality follows from the fundamental theorem of calculus.

Thus:
\begin{align} 
& \lim_{\epsilon \rightarrow 0} \epsilon^{-d}  \sum_{b \in \{0, 1\}^d} (-1)^{\|b\|_1} \cdot f ( \langle a + \epsilon b, {\bf 1} \rangle  ) \notag \\
 = & ~ \lim_{\epsilon \rightarrow 0} \epsilon^{-d} \int_0^\epsilon \int_0^\epsilon \ldots \int_0^\epsilon f^{(d)}( \langle a+ x, {\bf 1}\rangle ) \d x_1 \ldots \d x_d \notag \\
 = & ~ 
f^{(d)} (\langle a, \bf{1} \rangle ) \notag
\end{align}
where the first equality follows from Eq.~\eqref{eq:simplified} and the last equality follows from the continuity of $f^{(d)}$ This completes the proof of Lemma~\ref{lem:converge}.
\end{proof}

\subsection{Main Result}\label{sec:poly_main}

In this section, we prove main result Theorem~\ref{thm:log_formal} using Lemma~\ref{lem:poly_imply} and Lemma~\ref{lem:zero_eig}.

\begin{theorem}[Formal statement of Theorem~\ref{thm:log}]\label{thm:log_formal}
For any positive integer $n \geq 2$, the function $f : \R \to \R$ preserves low rank matrices if and only if $f$ is a polynomial of degree less than $\lceil \log_2(n) \rceil$.
\end{theorem}

\begin{proof} 
Suppose $f$ is a Taylor expandable function. By Lemma~\ref{lem:poly_imply}, we can find $M:\mathbb{R}^d \rightarrow \mathbb{R}^{n \times n}$ and a Taylor expandable $\lambda_i^f:\mathbb{R}^{\log n + 1} \rightarrow \R$ such that the image of $M$ has rank $\leq \log n + 1$, and $\{\lambda_i^f(a)\}_{i \in [n]}$ is the full set of eigenvalues of $M(a)$. Further, if there exists $i \in [n]$ with function $\lambda_i^f = 0$, then $f$ is a degree $d \leq \log n + 1$ polynomial.

Now, suppose that $f$ is a function that transforms all rank $\log n +1$ matrices to rank $<n$ matrices. Then it must transform all matrices $M(a)$ to rank $< n$ matrices. By Lemma~\ref{lem:zero_eig}, it must follow that $\lambda_i^f = 0$ for some $i$. However, we just established via Lemma~\ref{lem:poly_imply} that if $\lambda_i^f = 0$, then $f$ is a degree $d \leq \log n +1$ polynomial. This completes the proof of Theorem~\ref{thm:log_formal}.
\end{proof}


\section{Transforming Manhattan to Euclidean}\label{sec:manhattan_transform12}

In this section, we prove Theorem~\ref{thm:formal_manhattan_transform_1_and_2}, which states that
functions $f$ that transform Manhattan distances to squared Euclidean
distances are Bernstein.  This section is organized as follows

\begin{itemize}
   \item In Section~\ref{sec:manhattan_transform12:converse},  we show that any function $f:\mathbb R_{\geq 0}\to\mathbb R_{\geq 0}$ transforming
Manhattan to squared Euclidean is increasing. This serves as a warm-up for the general result which involves higher difference operations.
    \item In Section~\ref{sec:manhattan_transform12:main}, we prove the main result of this section, Theorem~\ref{thm:formal_manhattan_transform_1_and_2}.
    \item In Section~\ref{sec:manhattan_transform12:boundcont}, Lemma~\ref{lem:bound-cont} shows $f$ must be bounded and
continuous. This lemma is used in the proof of our main result.
\end{itemize}

\subsection{Useful Computations}\label{sec:manhattan_transform12:converse}

\begin{lemma}\label{lem:1deriv} 
If $f$ transforms Manhattan to squared Euclidean, then $f$ is increasing on $\mathbb R_+$. 
\end{lemma}

\begin{proof}
We fix $c>0$ and show $f'(c)\geq 0$. Consider $\chi:[d] \rightarrow
\{0, 1\}$ which transforms $1$ to $1$ and everything else to $0$. Let
$a_1 = \epsilon$ and $a_2, \ldots a_d = \frac{2c}{d}$. Here, $\epsilon$
is a constant which we will adjust later. 

The eigenvalue corresponding to $\chi$ (by Lemma~\ref{lem:fourier_formal}) is, by straightforward calculation:

\begin{align} \label{eq:binom}
\sum_{s = 0}^{d-1} \binom{d-1}{s} \left(f\left(\frac{2cs}{d}\right)-
    f\left(\frac{2cs}{d}+\epsilon\right)\right) 
\end{align} 

If we divide by $2^{d-1}$ and take $d$ to to infinity, the quantity in
Eq.~\eqref{eq:binom} becomes
\[ f(c) - f(c+\epsilon) \]
for continuous functions $f$. Indeed, nearly all of the probability mass in the binomial coefficients concentrates around $s=d/2$ by the law of large numbers and the limit follows from continuity of $f$ and the boundedness of $f$ on bounded sets established below in Lemma~\ref{lem:bound-cont}.\\

Applying Lemma~\ref{lem:euc-variant}, we see that if $f$ transforms Manhattan to squared Euclidean distances, then $f(c)- f(c+\epsilon)\leq 0$ for    any $\varepsilon>0$. This implies the desired result.
\end{proof}

\subsection{Main Results}\label{sec:manhattan_transform12:main}

The goal of this section is to prove Theorem~\ref{thm:formal_manhattan_transform_1_and_2}.
\begin{theorem} [Manhattan to squared Euclidean, formal version of part (1) $\Leftrightarrow$ part (3) of Theorem~\ref{thm:informal_manhattan_transform}]\label{thm:formal_manhattan_transform_1_and_2}
If $f$ transforms
Manhattan distances to squared Euclidean distances, it must be Bernstein. 
\end{theorem}

\begin{proof}
Fix a $k$-tuple $\epsilon=(\epsilon_1,\dots,\epsilon_k)$ of positive real numbers and define
\[ 
\Delta_{\epsilon}^k (f, t) := f(t) - \sum_{i_1\in [k]} f(t+\epsilon_{i_1})
    + \sum_{i_1<i_2\in [k]} f(t+\epsilon_{i_1}+\epsilon_{i+2})  +
    \ldots + (-1)^k f\left(t+\sum_{i=1}^k \epsilon_i\right). 
    \]

Consider $\chi$ that transforms $1, 2, \ldots k$ to $1$ and everything else
    to $0$. 
    Let $a_i= \epsilon_i$ for $ i \in [k]  $ and $a_{k+1} \ldots a_d =
    \frac{2c}{d}$ where $c$, $k$ and $\epsilon$ are fixed.
    
The eigenvalue corresponding to $\chi$ is, by direct
calculation using Lemma~\ref{lem:fourier_formal}:
\begin{align}\label{eq:k_binom}
\lambda_{\chi}=\sum_{s = 0}^{d-k} \binom{d-k}{s} \Delta_\epsilon^k(f, 2sc/d).
\end{align}
Eq.~\eqref{eq:k_binom} is the $d$-dimensional analog of Eq.~\eqref{eq:binom}, and this eigenvalue must satisfy $\lambda_{\chi}\leq 0$ by Lemma~\ref{lem:euc-variant}.  
Dividing by $2^{d-k}$ and taking $d$ to infinity, we obtain:
\begin{align*}
\Delta_{\epsilon}^k(f, c)\leq 0. 
\end{align*}

This is because again the probability mass in the binomial coefficients
in Eq.~\eqref{eq:k_binom} concentrates around the
$s=d/2$ coefficient, where we use continuity and boundedness of
$f$ for any compact set (guaranteed by Lemma~\ref{lem:bound-cont}). By Proposition~\ref{prop:bern} this implies $f$ is Bernstein (Definition~\ref{def:bernstein}) since $k,c$ were arbitrary. This completes the proof.
  \end{proof}

\subsection{Function Should be Bounded}\label{sec:manhattan_transform12:boundcont}
  The goal of this section is to prove Lemma~\ref{lem:bound-cont}.

  \begin{lemma} \label{lem:bound-cont} Any function $f:\mathbb R_{\geq 0}\to\mathbb R_{\geq 0}$ that transforms
  Manhattan to squared Euclidean is bounded on bounded sets and
  continuous on $(0, \infty)$.
  \end{lemma} 
  \begin{proof} 
  By the triangle inequality, $f(x) \le f(1/2)+f(1/2)$ for all $0 \le x \le 1,$ so $f$ is bounded on $[0,1]$.
  By scaling, from now on we assume $f$ is bounded by $1$ on $[0, 1]$.

  Now, we show $f$ is continuous on $(0, 1)$. Suppose there is a discontinuity at some point $0 < p < 1$. This means that there exists some $\varepsilon$ such that for all $\delta > 0,$ there are $a, b \in [p-\delta, p+\delta]$ such that $f(a)-f(b) \ge \varepsilon.$ Since $f(x) \le 1$ for all $x \in [0, 1],$ this means that for all $\delta < \min \{ p, 1-p\},$ we have that $\frac{f(a)}{f(b)} > 1+\varepsilon.$
  
  Now, fix some $\varepsilon$ satisfying the above, and some $n = 2k$. Consider points $x_1, \dots, x_n$ partitioned into sets $A = x_1, \ldots, x_k$ and $B= x_{k+1}, \ldots, x_n$. For some small $\delta$ that we will choose later, pick $a, b \in [p-\delta, p+\delta]$ such that $\frac{f(a)}{f(b)} > 1+\varepsilon,$ and define the metric
  \[d(x_i, x_j) :=
    \begin{cases}
    0 & i = j \\
    a & i,j\in A, i \neq j\\
    b & i \text{ or } j \text{ is in } B, i \neq j\\
    \end{cases}
  \]
  
    Now apply $f$: this gives us some metric $d'(x_i, x_j)$ such that
  \[d'(x_i, x_j) :=
    \begin{cases}
    0 & i = j \\
    f(a) & i,j\in A, i \neq j\\
    f(b) & i \text{ or } j \text{ is in } B, i \neq j\\
    \end{cases}
  \]
  
  We show that matrix $D'_{i, j} := d'(x_i, x_j)$ is not negative type if $n$ is sufficiently large (as a function of $\varepsilon$).
  Consider the vector 
  \begin{align*}
   v = (1, 1, \ldots 1, -1, -1, \ldots -1) 
  \end{align*}
  with the first $k$ coordinates are ones and the last $k$ coordinates are negative ones. This is orthogonal to the all ones
  vector, but 
  \begin{align*}
    v^{\top} D' v 
    = & ~ k(k-1)f(a) - 2k^2f(b) + k(k-1) f(b) \\
    = & ~ k(k-1) f(a) - k(k+1) f(b).
 \end{align*}
  Since $\frac{f(a)}{f(b)} > 1+\varepsilon,$ if we choose $n > 100/\varepsilon,$ we will have that
  \begin{align*}
  k(k-1) \cdot f(a) - k(k+1) \cdot f(b) > 0.
  \end{align*}
  
  Therefore, by Lemma \ref{lem:euc}, $d'$ does not embed into $\ell_2^2,$ Squared Euclidean space.
  
  However, we show that if $\delta$ is sufficiently small (in terms of $n, p$), then $d(x_i, x_j)$ is embeddable into $\ell_1$. First note that the metric $d_1(i, j)$ which equals $0$ if $i = j$ and $c$ for some constant $c > 0$ is embeddable into $\ell_1,$ by transforming $i$ to $x_i = \frac{c}{2} \cdot e_i$ for all $i$, where $e_i$ is the $i$th unit vector. Likewise, the metric $d_{k, \ell}(i, j)$ which equals $0$ if $i = j$ or if $i = k, j = \ell$ or $i = \ell, j = k$ and $c$ otherwise is also embeddable into $\ell_1,$ by transforming $i$ to $x_i = \frac{c}{2} \cdot e_i,$ except $\ell$ which is sent to $x_\ell = x_k = \frac{c}{2} \cdot e_k$. Now, it is trivial to see that by adding a finite number of these metrics, we still get a metric that is embeddable into $\ell_1$. 
  
  But, if $\frac{a}{b} \in \left[1 - \frac{1}{10 n^2}, 1 + \frac{1}{10 n^2}\right],$ then any metric such that $d(i, j) \in \{a, b\}$  
  for all $a, b$ can be written as some positive finite combination of $d_1$ and $d_{k, \ell}$ over all $1 \le k < \ell \le n$.
  
  Therefore, if $f$ is discontinuous at $p$, we can set $n = \frac{100}{\varepsilon}$, $\delta = \frac{\min(p, 1-p)}{100 n^2}$, and the metric on $x_1, \dots, x_n$ as defined previously. We will have that 
  \begin{align*}
  \frac{a}{b} \in \left[1 - \frac{1}{10 n^2}, 1 + \frac{1}{10 n^2}\right] 
  \end{align*}
  whereas $\frac{f(a)}{f(b)} > 1+\varepsilon,$ which means that while $d$ is embeddable into $\ell_1,$ $d' = f(d)$ is not embeddable into $\ell_2^2$. Thus, if $f$ is discontinuous at $p$, we have that $f$ cannot transform Manhattan Distances to Squared Euclidean distances.
  
  By scaling the $x$-axis, we have that $f$ is bounded on any interval $[0, a]$ and that $f$ is continuous at all $x > 0$.
  \end{proof}

\section{Transforming Manhattan to Manhattan}\label{sec:manhattan_transform23}

This section is organized as follows:
\begin{itemize}
    \item Section~\ref{sec:manhattan_transform23:tools} provides some useful tools that are related to $\ell_1$ distance, $\ell_2$ distance and Hadamard transform.
    \item In Section~\ref{sec:manhattan_transform23:main}, we prove Theorem~\ref{thm:formal_manhattan_transform_2_and_3} which is the main result.
    \item Section~\ref{sec:manhattan_transform23:discussion} provide some discussions.
\end{itemize}

\subsection{Useful Tools}\label{sec:manhattan_transform23:tools}
Suppose $f$ transforms Manhattan distance to squared Euclidean distance. By definition, $f$ satisfies the following: for any $n$ and any $x_1, \ldots x_n \in
(\mathbb{R}^{\mathbb{N}}, \ell_1)$, 
there exist $p_1, \ldots p_n \in (\mathbb{R}^{\mathbb{N}}, \ell_2)$ such that $f(\|x_i -
    x_j\|_1) = \|p_i - p_j\|_2^2$.  We can assume
without loss of generality that points $x_1, x_2, \ldots x_n$ are distinct corners of a $d$ dimensional
hyperrectangle (Definition~\ref{def:hyperrectangle}), and $n = 2^d$. This is because any point set in $\ell_1$ can be
embedded isometrically into $\ell_1$ on corners of a hyperrectangle
(Lemma~\ref{lem:l1-hyperrectangle}).

\begin{lemma}\label{lem:had}
Let $f:\mathbb{R}\rightarrow \mathbb{R}$. If $x_1, \ldots x_{2^d}$ are corners of a hyperrectangle (Definition~\ref{def:hyperrectangle}),
  the matrix $D$ where $D_{i,j} = f(\|x_i-x_j\|_1)$  must have
 eigenvectors which are the columns of $H_d$, where $H_d$ is the
  Hadamard matrix of size $2^d$ by $2^d$.
\end{lemma}
 \begin{proof} 
   This follows from Lemma~\ref{lem:key}.
 We note that this lemma does not rely on any assumptions on $f$.
 \end{proof}

 \begin{lemma}\label{lem:eig3} Let $D$ be the matrix where $D_{i,j} =
   f(\|x_i - x_j\|_1)$, and let $M:= -\frac{1}{2} \Pi D \Pi$. Then $M$ has eigenvectors $H_d$.
\end{lemma}
\begin{proof} This follows from Lemma~\ref{lem:had} and the definition
of $M$. It is critically important that the columns of $H_d$ are
orthogonal to the all ones vector (with the exception of the all ones
  column in $H_d$).
\end{proof}
\begin{lemma}\label{lem:p}
 Let $M = H_d \Sigma H_d$ be an eigendecomposition of
 $M$, where $M$ is defined as in Lemma~\ref{lem:eig3}. If $f$ transforms
  $\ell_1$ to $\ell_2^2$,  then $\Sigma$ has entirely non-negative
  entries.

For each $i$, we use  $p_i$ to denote the $i$-th column of $P = \sqrt{\Sigma} H_d$, we have
 $\langle p_i, p_j \rangle = M_{i,j}$ and $f(\|x_i - x_j\|_1) =
 \|p_i-p_j\|_2^2$.
 \end{lemma}
 \begin{proof} This follows from Lemma~\ref{lem:emb} and
   Lemma~\ref{lem:eig3}.
 \end{proof}

\subsection{Main Result}\label{sec:manhattan_transform23:main}

The goal of this section is to prove Theorem~\ref{thm:formal_manhattan_transform_2_and_3}.
\begin{theorem} [Manhattan to squared Euclidean, formal version of part (2) $\Leftrightarrow$ part (3) of Theorem~\ref{thm:informal_manhattan_transform}]\label{thm:formal_manhattan_transform_2_and_3} Any function that transforms Manhattan distances to squared
Euclidean distances must transform Manhattan distances to Manhattan distances, and vice versa. 
\end{theorem}

\begin{proof}  
  Let $p_i$ be defined as in Lemma~\ref{lem:p}.  By construction, the
  vectors $p_i$ are a subset of the corners of a $2^d$-dimensional hyperrectangle, with side
 lengths $\sqrt{\Sigma_{i,i}}$. Thus, the
 pairwise squared Euclidean distances between $p_i$ are isometrically
 embeddable into $\ell_1$ by Lemma~\ref{lem:l2-hyperrectangle}. In other words, $f(\|x_i - x_j\|_1) = \|p_i -
 p_j \|_2^2 = \|q_i -
 q_j\|_1$ for some $q_i \in \ell_1$ for all $i, j$. This shows that any $f$ that transforms $\ell_1$ to
 $\ell_2^2$ transforms $\ell_1$ to $\ell_1$ as
 desired.
 \end{proof}

Note that for any $x_i$, the vectors $q_i$ are finite dimensional and
can be explicitly written down in closed form.

\subsection{Discussion and Extensions}\label{sec:manhattan_transform23:discussion}
In our proof of Theorem~\ref{thm:formal_manhattan_transform_2_and_3}, we exploited that our points $x_1, \ldots x_n$ are points
in a hyperrectangle, which has a vertex transitive group symmetry. Similar theories
can be generated when the point set lives on any object with a
vertex-transitive group symmetry, and the distance measure between
points is some function of the Euclidean distance. Such objects include higher dimensional
platonic solids, spheres, equilateral triangular prisms, and more.

We remark that the group symmetry must be vertex-transitive to ensure the matrix $D$ in Lemma~\ref{lem:had} has an eigenvector equal to the all ones vector. If
this were not the case, Lemma~\ref{lem:eig3} would no longer hold.


\section{Positive Definite Manhattan Kernels}\label{sec:bern}

The section is organized as follows:
\begin{itemize}
    \item In Section~\ref{sec:bern:tool}, we state a useful tool.
    \item In Section~\ref{sec:bern:main}, we present our main result. This result classifies all positive definite Manhattan kernels (Definition~\ref{def:manhattan_kernel}), and is a formal restatement of Theorem~\ref{thm:informal_kernel_manhattan}.
\end{itemize}

\subsection{A Useful Tool}\label{sec:bern:tool} 

First, we prove the following lemma.
\begin{lemma}

\label{lem:monotonepositive}

If $f$ is a positive definite Manhattan kernel (Definition~\ref{def:manhattan_kernel}), then $f(t)\geq 0$ for all $t\geq 0$.

\end{lemma}

\begin{proof}

Let ${\cal X}$ denote metric space $(\R^N, \ell_1)$.  For any $N \geq 0$ we consider the points $x_i=\frac{t}{2}e_i\in {\cal X}$ for $i\in [N]$ where $e_i=(0,\dots,0,1,0,\dots,0)$ is a standard basis vector, so that $\|x_i-x_j\|_1=t$ for any $i\neq j$. Since the matrix of values $(f(\|x_i - x_j\|_1)_{i,j\in [N]}$ must be positive semidefinite, the sum of all its entries must be positive, hence:
\begin{align*}
N(f(0)+(N-1)f(t))\geq 0.
\end{align*}
The above equation implies the following:
\begin{align*}
\frac{f(0)}{N-1}+f(t)\geq 0
\end{align*}
for all integer $N \geq 0$ and real $t \geq 0$. 

Since $N$ can be arbitrarily large, therefore we conclude $f(t)\geq 0$ as claimed.

\end{proof}

\subsection{Main Result}\label{sec:bern:main}

The goal of this section is to prove Theorem~\ref{thm:formal_kernel_manhattan}.
\begin{theorem}[Formal statement of Theorem~\ref{thm:informal_kernel_manhattan}] \label{thm:formal_kernel_manhattan} 
  $f:\mathbb R_{\geq 0}\to \mathbb R$
  is a positive definite Manhattan kernel (Definition~\ref{def:manhattan_kernel}) if and only if
  $f(x)$ is completely monotone (Definition~\ref{def:cm}). 
\end{theorem}
\begin{proof}

First, we prove that if $f$ is a positive definite Manhattan kernel, then $f$ must be completely monotone.  The converse direction is previously known, and is a consequence of Lemma~\ref{lem:l1-iso} and Theorem 3 of~\cite{sow01}\footnote{Theorem 3 of ~\cite{sow01} is a modern restatement of Schoenberg's work in~\cite{s42}}.

Suppose that $f$ is a positive definite Manhattan kernel (Definition~\ref{def:manhattan_kernel}). Cauchy-Schwarz easily implies that $f(t)\leq f(0)$ for all $t$, so $f$ is bounded.. Now, if $x_1,\dots, x_n$ correspond to $y_1,\dots,y_n$ then
\begin{align*}
f(\|x_i-x_j\|_1)
= & ~ \langle y_i,y_j\rangle \\
= & ~ f(0)-\frac{1}{2} \|y_i-y_j\|_2^2. 
\end{align*}

Therefore $2(f(0)-f(t))$ (equivalently, $f(0)-f(t)$) sends Manhattan distances to squared Euclidean distances. Therefore $f(0)-f(t)$ is Bernstein (Definition~\ref{def:bernstein}), by Theorem~\ref{thm:informal_manhattan_transform}. Combining with Lemma~\ref{lem:monotonepositive} we conclude that $f$ must be completely monotone (Definition~\ref{def:cm}). 

\end{proof}

\section{Representation Theory of the Real Hyperrectangles}\label{sec:key}
In this section, we prove Lemma~\ref{lem:fourier_formal}, the formal version of Lemma~\ref{lem:fourier_informal}. This lemma uses representation theoretic ideas to compute the eigenvalues of matrices arising from the real hyperrectangle. We introduce Lemma~\ref{lem:box-int}, which expresses these same eigenvalues in terms of integrals. This integral formulation is useful for proving Theorem~\ref{thm:log_formal}.

\subsection{Useful Tools}

\begin{lemma}~\label{lem:key} Let $g:(\mathbb{R}^d \times \mathbb{R}^d)
  \rightarrow \mathbb{R}$ such that $g(x,y)$ is invariant under axis
  reflection.  Consider a $d$-dimensional hyperrectangle with
corners $x_1, \ldots x_{2^d}$.  
  Let $D$ be a $2^d$ by $2^d$ matrix such that $D_{ij} = g(x_i, x_j)$.
  Then there is an eigendecomposition of $D$ into $H_d \Sigma H_d$ where $\Sigma$ is a
diagonal matrix.
\end{lemma}
\begin{proof} This lemma can be proven directly via computation.
However, it is more instructive to view this through the representation
theoretic lens. We note that $D$ has the property that for any
permutation matrix $\sigma$ corresponding to a reflection about one of
the hyperrectangle's axes, we have $\sigma D = D\sigma$. Schur's lemma from
representation theory (see Lemma~\ref{lem:known-abelian}) states that $D$
  and all $\sigma$ in the reflectional symmetry group of the hyperrectangle have a common set of eigenvectors. It is straightforward
to verify that the only common set of eigenvectors for all $\sigma$ is
the columns of the Hadamard matrix, and thus $D$ must have the columns
of $H_d$ as its eigenvectors.
\end{proof}
We note that variants of this lemma are used to prove Delsarte's linear programming
bound in error correcting codes~\cite{delsarte, ODonnell14}.
\subsection{Main Result}
\begin{lemma}[Representation theory of the real hyperrectangle, formal version of Lemma~\ref{lem:fourier_informal}]~\label{lem:fourier_formal}
Consider a $d$-dimensional hyperrectangle (Definition~\ref{def:hyperrectangle}) parameterized by $a_1, \ldots a_d > 0$. Enumerate the vertices in lexicographical ordering as $p_1, \ldots p_{2^d}$.
 
For any $f: \R \to \R$, let $D$ be the $2^d$ by $2^d$ matrix given by $D_{i,j} =f(\| p_i - p_j \|_1)$. Then:
 \begin{enumerate}
     \item $\Sigma := H_d D H_d$ is a diagonal matrix whose entries are the eigenvalues of $D$ multiplied by $2^d$, and $D = 4^{-d} \cdot H_d \Sigma H_d $.
     \item Let $\chi: [d] \rightarrow \{0, 1\}$. Let $k$ equal the integer corresponding to transforming $\chi$ (written as a $d$ dimensional binary vector) into an integer via binary conversion. For each $\chi$, there is an eigenvector of $D$ equal to the $k$-th column of Hadamard matrix $H_d$, and its associated eigenvalue is:
     \begin{align}\label{eq:eig0}
\sum_{T \subseteq [d]} (-1)^{\sum_{t \in T} \chi(t)} f\left(\sum_{t\in T}a_t\right).
\end{align}
\end{enumerate}
\end{lemma}

The second part of this theorem on its surface differs from that in Lemma~\ref{lem:fourier_informal}, but the statements are in fact identical via straightforward computation.
\begin{proof} By Lemma~\ref{lem:key}, we know that the Hadamard matrix columns are eigenvectors of the matrix $D$. The result follows by direct computation, noting that the formula in Eq.~\eqref{eq:eig0} is the $d$ dimensional analog of
Eq.~\eqref{eq:eigval}, and can be derived in the same way. \end{proof}

We now give an alternate formulation of the eigenvalues in
Lemma~\ref{lem:fourier_formal}. This lemma is of independent interest.
\begin{lemma} \label{lem:box-int}
Given a box with side lengths $a_1, \ldots a_d$, each eigenvalue
analogous to those in Eq.~\eqref{eq:eigval} corresponds to a function
$\chi: [d] \rightarrow \{0, 1\}$.  Let $Q = \{q_1, \ldots q_k\}$ be the
full set of values on which $\chi$ is $1$.
Then the Eigenvalues in Eq.~\eqref{eq:eig0} equal:
\begin{align*}
  \sum_{T \subseteq [d] \setminus Q}
  \int_{\sum_{t \in T} a_t}^{a_{q_1}+\sum_{t \in T} a_t} \ldots
  \int_{\sum_{t \in T} a_t}^{a_{q_k}+\sum_{t \in T} a_k}
  (-1)^k\frac{\d^k f}{\d x^k} \Big( \sum_{q \in Q} s_q \Big) \d s_1 \ldots \d s_k.
\end{align*}
\end{lemma}
\begin{proof}
The proof is identical to that of Eq.~\eqref{eq:int}, but for $d$ dimensions. It follows directly from Lemma~\ref{lem:fourier_formal} combined with the fundamental theorem of calculus.
\end{proof}

\end{document}